\documentclass[conference]{IEEEtran}
\IEEEoverridecommandlockouts
\usepackage{cite}
\usepackage{amsmath,amssymb,amsfonts}
\usepackage{algorithmic}
\usepackage{graphicx}
\usepackage{textcomp}
\usepackage{xcolor}
\usepackage{url}
\usepackage{hyperref}
\usepackage{xurl}

\usepackage{tikz}
\usepackage{textcomp}

\newcommand\copyrighttext{%
  \footnotesize \textcopyright 2026 IEEE. Personal use of this material is permitted.  Permission from IEEE must be obtained for all other uses, in any current or future media, including reprinting/republishing this material for advertising or promotional purposes, creating new collective works, for resale or redistribution to servers or lists, or reuse of any copyrighted component of this work in other works. 

  Accepted for publication in ICSA-C 2026, 23rd IEEE International Conference on Software Architecture Companion Proceedings.}
\newcommand{\copyrightnotice}{%
\begin{tikzpicture}[remember picture,overlay]
\node[anchor=south,yshift=10pt] at (current page.south) {\fbox{\parbox{\dimexpr\textwidth-\fboxsep-\fboxrule\relax}{\copyrighttext}}};
\end{tikzpicture}%
}

\begin{document}

\title{The Need for a Green ICT Reference Framework\thanks{Marco Aiello and Ilche Georgievski are with IAAS, University of Stuttgart, Germany; Mina Alipour and Mahyar Tourchi Moghaddam are with University of Southern Denmark; Antonio Brogi is with University of Pisa, Italy; Rafael Capilla is with Universidad Rey Juan Carlos, Spain; Lidia Fuentes is with Universidad de Málaga, Spain; Gabriele Gianini is with the University of Milano-Bicocca, Italy; Monica Vitali is with Politecnico di Milano, Italy; Sebastian Werner is with the TU Berlin, Germany}}

\author{\IEEEauthorblockN{Marco Aiello, Mina Alipour, Antonio Brogi, Rafael Capilla, Lidia Fuentes, Ilche Georgievski, \\Gabriele Gianini, Mahyar Tourchi Moghaddam, Monica Vitali, Sebastian Werner}
\IEEEauthorblockA{\textit{Green ICT Working Group, Informatics Europe}, 
Zurich, Switzerland\\
\url{https://www.informatics-europe.org/society/green-ict.html}
    }
}

\maketitle

\copyrightnotice

\begin{abstract}
The sustainability impacts of ICT systems are difficult to assess and govern due to structural complexity, fragmented measurement practices, and unclear responsibilities across system layers. 
We argue that these challenges cannot be addressed solely by metrics and motivate the need for a shared Green ICT reference framework that integrates sustainability across multiple perspectives and domains, lifecycle phases, and governance contexts. We present an initial framework developed within the Informatics Europe Green ICT Working Group as a first step towards a comprehensive reference framework.
\end{abstract}


\section{Introduction}
\label{sec:intro}

The environmental impact of Information and Communication Technologies (ICT) has reached
a level that demands primary consideration.
Data centers, communication networks, cloud platforms, software systems, 
and increasingly AI-driven workloads collectively form a global digital infrastructure whose energy demand continues to grow~\cite{bogmans2025power}.
At the same time, ICT is widely promoted as a key enabler of sustainability in various domains, including smart energy grids, mobility systems, climate modelling, and resource optimisation. {\it Green ICT} has emerged as a prominent term in research agendas, policy documents, and corporate strategies, pushing organisations to generate annual sustainability reports to prove their compliance with green regulations. Yet, despite this visibility, Green ICT practice often remains superficial in how sustainability concerns are operationalised: sustainability is frequently addressed through disconnected practices at different levels, ranging from narrow-focused metrics and high-level dashboards to marketing-driven commitments, rather than being embedded into the core of system design and engineering.
This limitation reflects a broader challenge identified in the literature: sustainability is often not explicitly treated as a systemic property of software and ICT systems, but rather addressed implicitly or in isolation, despite calls to consider it across architectural layers and lifecycle stages \cite{becker2015sustainability, lago2015framing}.
In the worst cases, delegating computational components to external services to avoid accountability for emissions becomes a form of greenwashing~\cite{famularo2023corporate}.
Despite existing work, ranging from sustainability metrics and energy-aware programming techniques to organisational guidelines and sector-specific standards
\cite{10.1007/978-3-319-44711-7_4}\cite{10.1145/3728638}\cite{din_iso2025}, current approaches are fragmented, largely layer-specific, and insufficiently integrated across architectural levels, lifecycles, and governance, leaving a clear gap for a unifying Green ICT reference framework\cite{din_iso2025}.
The Informatics Europe Green ICT Working Group is developing such a framework that we sketch in this position paper.

\begin{figure*}[t]
  \centering
  \includegraphics[width=0.80\textwidth]{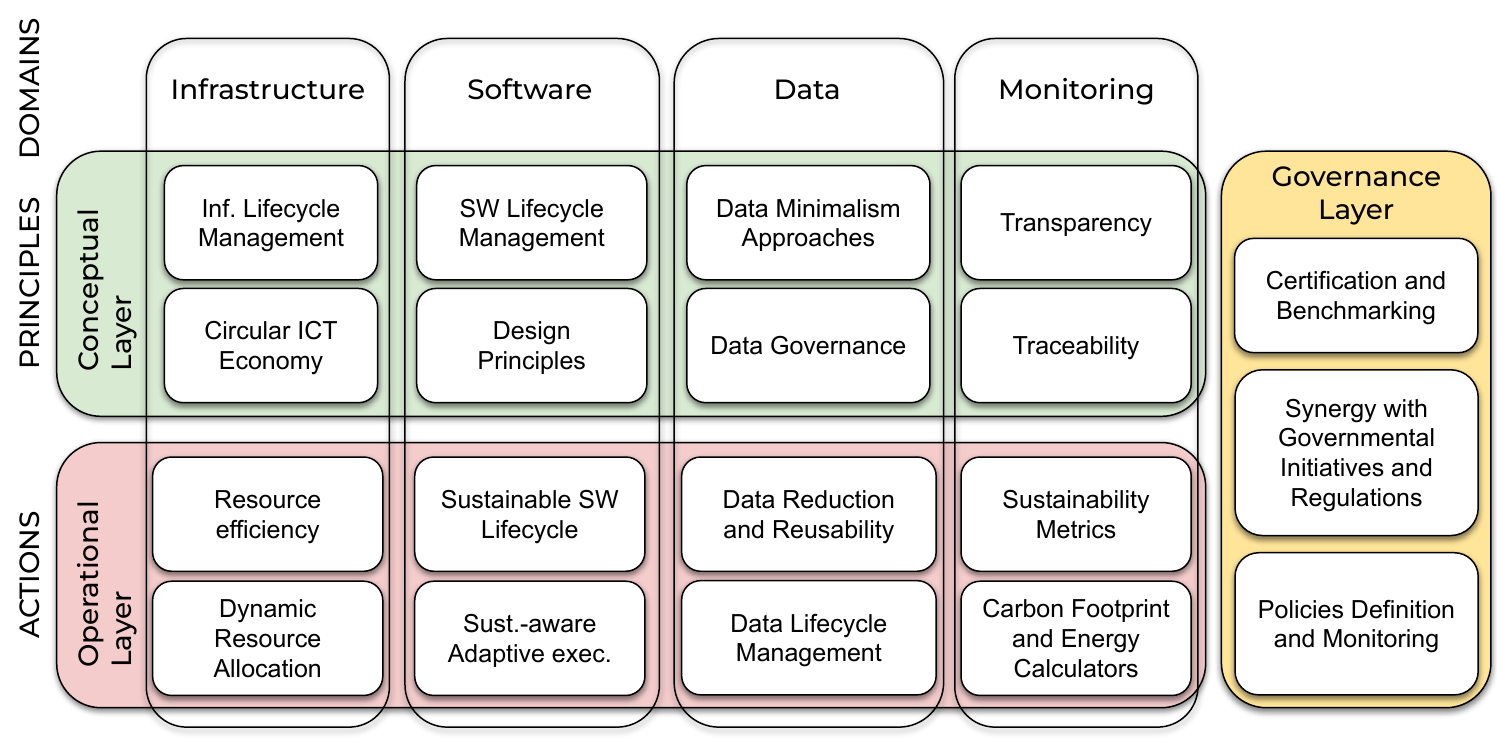}
  \caption{A General Framework for Green ICT}\label{fig:framework}
\end{figure*}

\section{Challenges of Modern ICT}
ICT systems face a set of structural, methodological, and governance challenges hindering a rigorous and meaningful sustainability assessment and accountability. 
A defining characteristic of contemporary ICT systems is that their architectures must reconcile demanding operational environments (e.g., cloud platforms, distributed data) with stakeholder and governance concerns (e.g., regulation, organisational strategy), driving the need for cross-domain mechanisms and systems architectures. Applications run on top of complex software stacks that themselves rely on virtualisation platforms, networks, distributed and remote servers, and cooling infrastructures. 
From a sustainability perspective, these characteristics imply that energy consumption and carbon emissions are emergent properties of the entire stack, rather than attributes of individual components. Additionally, sustainability is in a constant trade-off with performance and Quality of Service (QoS), since some architectural choices like replication or scaling affect performance while increasing the overall carbon footprint. The resulting lack of clear attribution, measurement, transparency and responsibility across layers constitutes a central challenge for sustainability assessment in ICT systems.

Assessing the environmental impact of ICT is inherently complex. Numerous sustainability metrics have been proposed both at the infrastructure level (e.g., Power Usage Effectiveness (PUE) and energy intensity~\cite{7921551}) and at higher levels (e.g., estimates derived from CPU utilisation, FLOPs per watt, or per-request energy consumption~\cite{10.1145/3728638}). For AI workloads, specialised tools aim to quantify the carbon footprint of training and inference~\cite{luccioni2023countingcarbonsurveyfactors}. Each metric embeds assumptions regarding system boundaries, workload characteristics, and energy sources. As measurements are aggregated across layers, uncertainties compound, leading to often unreliable and difficult to compare results. These metrics are ultimately interpreted by decision-makers, introducing additional ambiguity that extends beyond purely technical considerations.
Architectural complexity, fragmented metrics, and governance gaps can hide unsustainable ICT components, dilute emissions' responsibilities, and enable forms of greenwashing.  

How resource‑efficient software should be addressed across multiple levels—including software infrastructures, architectures, applications, and hardware environments—is also a matter of growing importance for standardization bodies. Notably, a recent recommendation emerging from an ISO open consultation on \textit{Measuring and Mitigating Carbon and Heat Emissions from AI Coding and Development Processes} emphasized the need to define a reference architecture model capable of more systematically identifying the relevant layers and their interdependencies\cite{din_iso2025}. Such a model would support a clearer conceptualization of the software stack and contribute to advancing the broader objective of improving software resource efficiency.

\section{The Green ICT Reference Framework}

Although metrics address isolated aspects of sustainability, they do not offer the conceptual structure required to align underlying assumptions, define system boundaries, allocate responsibilities, or assess system-wide impacts. A coherent reference framework that integrates sustainability concerns across ICT layers and domains is lacking\cite{din_iso2025}. The proposed framework is presented in Figure~\ref{fig:framework}. It was developed through the collective contributions of the Informatics Europe working group on Green ICT. The framework distinguishes between four domains (Infrastructure, Software, Data, and Monitoring), each defining key concepts that guide sustainability-aware design and assessment, applied to three main layers (i.e. Conceptual, Operational, and Governance).

\paragraph{Conceptual layer} The conceptual layer captures the guiding principles that shape sustainable ICT systems and provides the conceptual foundation for the operational layer. In the infrastructure domain, the conceptual layer defines \textit{Lifecycle Management} of hardware components, covering all stages from production to end-of-life, and introduces \textit{Circular ICT Economy} principles, promoting reuse, refurbishment, and recycling to minimise environmental impact\cite{cao2022toward}.
In the software domain, it establishes \textit{SW Lifecycle Management} as a core principle, encompassing all phases from design to decommissioning, and defines \textit{Design Principles} that incorporate sustainability as a primary quality attribute in software systems\cite{10.1145/3728638}.
In the data domain, the focus is on \textit{Data Minimalism Approaches}, encouraging the collection and retention of only necessary data\cite{vitale2018hoarding}, and on \textit{Data Governance}, which integrates sustainability considerations across the data lifecycle, including compliance, quality, privacy, and security\cite{otto2011data}.
Finally, the monitoring domain defines \textit{Transparency} and \textit{Traceability} as key principles\cite{hankendi2025transparency}, enabling the observation, measurement, and reporting of sustainability-related aspects across infrastructure, software, and data, and supporting informed decision-making through accessible and reliable information. 

\paragraph{Operational layer} The operational layer translates these principles into concrete actions and implementations. At the infrastructure level, this includes \textit{Resource Efficiency}, focusing on reducing energy and material consumption during operation, and \textit{Dynamic Resource Allocation}, enabling adaptive use of computational and storage resources based on demand and sustainability criteria\cite{bharany2022systematic}. 

In the software domain, the design principles and software lifecycle management from the conceptual layer are specialized into a \textit{Sustainable Software Lifecycle} that integrates sustainability considerations into development, deployment, and maintenance practices, and \textit{Sustainability-aware Adaptive Execution}\cite{vitali2025adaptive}, allowing software systems to dynamically adjust their behaviour in response to environmental or energy-related constraints and sustainability qualities defined in the \textit{design principles box}. The data domain emphasises \textit{Data Reduction and Reusability}, minimizing redundant data processing and promoting data reuse across applications, together with \textit{Data Lifecycle Management}, that aims at ensuring efficient handling of data from ingestion to deletion with reduced environmental impact\cite{schneider2023reuse}. Finally, the monitoring domain includes \textit{Sustainability Metrics}, aimed at providing quantifiable measures of energy consumption and environmental impact, as well as the use of \textit{Carbon Footprint and Energy Calculator} tools supporting the estimation and analysis of ICT-related emissions\cite{shao2022review} and energy consumption.

\paragraph{Governance layer} 
The governance layer complements the conceptual and operational layers by providing a transversal perspective that connects the framework with external standards, policies, and compliance with national or European regulation (e,g. detection of greenwashing practices). By acting as an interface between the framework and the broader regulatory and standardisation ecosystem, the governance layer enables the integration of external constraints and best practices, while supporting accountability and continuous alignment. This layer encompasses \textit{Certification and Benchmarking}, enabling the assessment and comparison of ICT systems against established sustainability standards and reference frameworks. It also supports \textit{Synergy with Governmental Initiatives and Regulations}, ensuring that ICT practices are aligned with evolving legal requirements and policy objectives. Finally, it supports \textit{Policies Definition and Monitoring}, providing mechanisms to define, enforce, and evaluate organisational sustainability strategies over time.

\section{Concluding Remarks and Future work}
\label{sec:conclusions}

Green ICT is hard to achieve, not because sustainability is incompatible with digital systems, but because modern ICT is inherently complex, layered, and distributed. In such environments, precision is limited, metrics are partial, and good intentions are easily diluted. We argue that, in the absence of a shared reference framework, Green ICT will remain fragmented, difficult to compare across contexts, and susceptible to greenwashing. While no framework can eliminate uncertainty, it can render sustainability considerations explicit, systematically structured, and subject to accountability. We therefore initiate the development of a general Green ICT reference framework and present its first outline here.
Once defined in sufficient detail, such a framework can serve as a foundation for multiple stakeholders. In education, sustainability can be the foundation for curricula that integrate sustainability into software engineering, systems architecture, and AI development. For policymakers, the framework enables more informed and timely regulation, while for industry, it provides a basis for defining accountable Green ICT strategies. 

Future work will focus on several dimensions. First, the initial outline of the framework will undergo refinement through a systematic literature review of each dimension, examining whether the concepts are precisely defined and maintain well-established mutual and collective relationships, and resolving any ambiguities through structured conceptual analysis. Second, the refined framework will be systematically validated by applying it to specific domains and case studies involving different infrastructure types, sectors, and geographies to test whether the boundaries, responsibilities, and metrics it proposes are complete and actionable in practice. Third, the refined framework would require operationalisation of its cross-layer dimensions, for example, by defining how emissions are attributed when a single workload spans multiple organisational and technical boundaries. Beyond this, it should be examined how the refined framework 
can be operationalised in education, policy, and practice.

\bibliographystyle{IEEEtran}
\bibliography{green}

@misc{luccioni2023countingcarbonsurveyfactors,
      title={{Counting Carbon: A Survey of Factors Influencing the Emissions of Machine Learning}}, 
      author={Alexandra Sasha Luccioni and Alex Hernandez-Garcia},
      year={2023},
      eprint={2302.08476},
      archivePrefix={arXiv},
      primaryClass={cs.LG},
      url={https://arxiv.org/abs/2302.08476}, 
}

@article{10.1145/3728638, 
author = {Danushi, Ornela and Forti, Stefano and Soldani, Jacopo},
title = {{Carbon-Efficient Software Design and Development: A Systematic Literature Review}}, 
year = {2025}, 
publisher = {ACM},
volume = {57}, 
number = {10}, 
journal = {ACM Comput. Surv.}
}

@InProceedings{10.1007/978-3-319-44711-7_4,
author="Verdecchia, Roberto
and Ricchiuti, Fabio
and Hankel, Albert
and Lago, Patricia
and Procaccianti, Giuseppe",
editor="Wohlgemuth, Volker
and Fuchs-Kittowski, Frank
and Wittmann, Jochen",
title={{Green ICT Research and Challenges}},
booktitle="Advances and New Trends in Environmental Informatics",
year="2017",
publisher="Springer",
pages="37--48"
}

@article{bogmans2025power,
  title={{Power hungry: How AI will drive energy demand}},
  author={Bogmans, Christian and Gomez-Gonzalez, Patricia and Ganpurev, Ganchimeg and Melina, Giovanni and Pescatori, Andrea and Thube, Sneha},
  journal={IMF Working Papers},
  volume={81},
  number={4},
  pages={2025},
  year={2025},
  publisher={International Monetary Fund}
}

@article{famularo2023corporate,
  title={Corporate social responsibility communication in the ICT sector: digital issues, greenwashing, and materiality}}

@article{vitali2025adaptive,
  title={Adaptive Green Cloud Applications: Balancing Emissions, Revenue, and User Experience through Approximate Computing},
  author={Vitali, Monica and Wiesner, Philipp and Kreutz, Kevin and Gandola, Roberto},
  journal={Future Generation Computer Systems},
  pages={108143},
  year={2025},
  publisher={Elsevier}
}

@ARTICLE{7921551,
  author={Reddy, V. Dinesh and Setz, Brian and Rao, G. Subrahmanya V. R. K. and Gangadharan, G. R. and Aiello, Marco},
  journal={IEEE Transactions on Sustainable Computing}, 
  title={Metrics for Sustainable Data Centers}, 
  year={2017},
  volume={2},
  number={3},
  pages={290-303},
  doi={10.1109/TSUSC.2017.2701883}}

@inproceedings{vitale2018hoarding,
  title={Hoarding and minimalism: Tendencies in digital data preservation},
  author={Vitale, Francesco and Janzen, Izabelle and McGrenere, Joanna},
  booktitle={Proceedings of the 2018 CHI Conference on Human Factors in Computing Systems},
  pages={1--12},
  year={2018}
}

@article{otto2011data,
  title={Data governance},
  author={Otto, Boris},
  journal={Business \& Information Systems Engineering},
  volume={3},
  number={4},
  pages={241--244},
  year={2011},
  publisher={Springer}
}

@article{cao2022toward,
  title={Toward a systematic survey for carbon neutral data centers},
  author={Cao, Zhiwei and Zhou, Xin and Hu, Han and Wang, Zhi and Wen, Yonggang},
  journal={IEEE Communications Surveys \& Tutorials},
  volume={24},
  number={2},
  pages={895--936},
  year={2022},
  publisher={IEEE}
}

@article{hankendi2025transparency,
  title={Why transparency matters for sustainable data centers and carbon-neutral artificial intelligence (AI)},
  author={Hankendi, Can and Coskun, Ayse K and Sovacool, Benjamin K},
  journal={iScience},
  volume={28},
  number={11},
  year={2025},
  publisher={Elsevier}
}

@article{shao2022review,
  title={A review of energy efficiency evaluation metrics for data centers},
  author={Shao, Xiaotong and Zhang, Zhongbin and Song, Ping and Feng, Yanzhen and Wang, Xiaolin},
  journal={Energy and buildings},
  volume={271},
  pages={112308},
  year={2022},
  publisher={Elsevier}
}

@article{bharany2022systematic,
  title={A systematic survey on energy-efficient techniques in sustainable cloud computing},
  author={Bharany, Salil and Sharma, Sandeep and Khalaf, Osamah Ibrahim and Abdulsahib, Ghaida Muttashar and Al Humaimeedy, Abeer S and Aldhyani, Theyazn HH and Maashi, Mashael and Alkahtani, Hasan},
  journal={Sustainability},
  volume={14},
  number={10},
  pages={6256},
  year={2022},
  publisher={MDPI}
}

@article{schneider2023reuse,
  title={Reuse, Reduce, Support: Design Principles for Green Data Mining: J. Schneider et al.: Reuse, Reduce, Support: Design Principles for Green Data Mining},
  author={Schneider, Johannes and Seidel, Stefan and Basalla, Marcus and vom Brocke, Jan},
  journal={Business \& Information Systems Engineering},
  volume={65},
  number={1},
  pages={65--83},
  year={2023},
  publisher={Springer}
}

@report{din_iso2025,
  title={ISO open consultation on Resource-Efficient Software: final report},
  author={DIN/ISO},
  institution={Deutsches Institut für Normung},
  year={2025},
  month={August},
  type={Proposed standard},
  doi          = {10.5281/zenodo.19088758},
  url          = {https://doi.org/10.5281/zenodo.19088758}
}

@inproceedings{becker2015sustainability,
  title={Sustainability design and software: The karlskrona manifesto},
  author={Becker, Christoph and Chitchyan, Ruzanna and Duboc, Leticia and Easterbrook, Steve and Penzenstadler, Birgit and Seyff, Norbert and Venters, Colin C},
  booktitle={2015 IEEE/ACM 37th IEEE International Conference on Software Engineering},
  volume={2},
  pages={467--476},
  year={2015},
  organization={IEEE}
}

@article{lago2015framing,
  title={Framing sustainability as a property of software quality},
  author={Lago, Patricia and Ko{\c{c}}ak, Sedef Akinli and Crnkovic, Ivica and Penzenstadler, Birgit},
  journal={Communications of the ACM},
  volume={58},
  number={10},
  pages={70--78},
  year={2015},
  publisher={ACM New York, NY, USA}
}

\end{document}